\RequirePackage{fix-cm}
\documentclass[smallextended]{svjour3}       % onecolumn (second format)
\smartqed  % flush right qed marks, e.g. at end of proof

\usepackage{graphicx}
\usepackage{subfigure}
\usepackage{amssymb}
\usepackage{color}
\usepackage{amsmath}
\usepackage{cite}
\usepackage{graphicx}
\usepackage{epstopdf}
\usepackage{cases}
 \journalname{Nonlinear Dynamics}
\begin{document}

\title{Families of exact solutions of a new extended $(2+1)$-dimensional Boussinesq equation  \thanks{Corresponding author, Jingsong He: hejingsong@nbu.edu.cn; jshe@ustc.edu.cn}
}
\subtitle{}


\author{Yulei Cao \and  Jingsong He \and Dumitru Mihalache
}


\institute{Yulei Cao, Jingsong He \at
              Department of Mathematics,  Ningbo University, Zhejiang 315211, P. R. China \\
              Tel.: +86 574-87600739\\
              Fax: +86 574-87600744\\
              \email{hejingsong@nbu.edu.cn; jshe@ustc.edu.cn}
\and Dumitru Mihalache \at Horia Hulubei National Institute for Physics and Nuclear Engineering, P.O.Box MG--6, Magurele, 077125, Romania}

\date{Received: date / Accepted: date}

\maketitle

\begin{abstract}
A new variant of the $(2+1)$-dimensional [$(2+1)d$] Boussinesq equation was recently introduced by J. Y. Zhu, arxiv:1704.02779v2, 2017; see
eq. (3). First, we derive in this paper the one-soliton solutions of both bright and dark types for the extended $(2+1)d$ Boussinesq equation
by using the traveling wave method.
Second, $N$-soliton, breather, and rational solutions  are obtained by using the Hirota bilinear method and the long wave limit. Nonsingular rational solutions of two types were obtained analytically, namely: (i) rogue-wave solutions having the form of W-shaped lines waves and (ii) lump-type solutions. Two generic types of  semi-rational solutions were also put forward. The obtained semi-rational solutions are as follows: (iii) a hybrid of a first-order lump and a bright one-soliton solution and (iv) a hybrid of a first-order lump and a first-order breather.

\end{abstract}

\section{Introduction}

During the last decades a wide variety of nonlinear evolution equation [NLEEs] have been used to model many interesting nonlinear phenomena in physics, chemistry, biology, and even in social sciences. The continuing fast growing research area of solitons or more properly solitary waves and their diverse applications in science and technology has attracted the interest of many research groups around the world since the pioneering work of Zabusky and Kruskal, published more than fifty years ago  \cite{zz1}, see also Refs. \cite{zz2,zz3,zz4,zz5,zz6,zz7,zz8,zz9,zz10,zz11,zz12}.  The observation of different types of solitons in a series of physical settings uncovered many interesting phenomena from both fundamental and applied points of view \cite{jp1,jp2,jp3,jp4,jp5,jp6,jp7,jp8,jp9}.

Due to the exact dynamical counterbalancing between nonlinear and dispersive effects, the solitary waves are wave packets that travel in nonlinear dispersive and/or diffractive media and retain their stable waveforms. The solitary waves in  $(1+1)$-dimensional [$(1+1)d$] settings have been researched extensively and are understood quite well, thus it was  necessary to study NLEEs in higher dimensions, e.g., in $(2+1)d$ and $(3+1)d$ physical settings. In particular, the exact solutions and their dynamics in the case of integrable $(2+1)d$ equations have been studied in detail, such as the Davey-Stewartson  [DS] equations \cite{OYY1,OYY2}, the Kadomtsev-Petviashvili-I [KPI] equation \cite{DDE1,DDE2}, the $(3+1)d$ KP equation  \cite{3DKP}, and other physically-relevant NLEEs, see, for example, Refs. \cite{MQ1,MQ2,MQ3,MQ4,MQ5,MQ6,MQ7,MQ8,MQ9,MQ10,MQ11,MQ12,MQ13,MQ14}.

More recently, the phenomenon of rogue waves [RWs] has become a hot research topic. These rogue (freak) waves ``appear from nowhere and disappear without a trace" \cite{AAM-1}. The concept of RWs has been extend beyond the common oceans RWs. During the past few years, many theoretical and experimental studies of RWs range from geophysics and hydrodynamics to oceanography, Bose-Einstein condensates \cite{BKA1,BKA2}, optics and photonics \cite{MBR1,MBR2}, plasma physics \cite{MMW1,MMW2}, superfluids \cite{an1}, and atmosphere physics \cite{an2}. Peregrine was the first one who obtained the fundamental RW solution of the generic nonlinear Schr\"{o}dinger [NLS] equation \cite{PHD1}. Recently, the higher-order RW solutions of the NLS-type equations  were studied in a series of articles by different methods \cite{OY1,OY2,OY3,OY4,OY44,OY5,OY6,OY7}. What is more, a hierarchy of other soliton equations have also been verified possessing different types of RW solutions \cite{PD1,PD2,PD3,PD4,PD5};
{for a recent review on rogue waves in scalar, vector, and multidimensional nonlinear systems, see Ref. \cite{review2017}.}

The dynamics of shallow water waves are governed by several kinds of NLEEs \cite{ab1,ab2,ab3,ab4,ab5,pa1,pa2,pa3,pa4}. Typically, the corresponding soliton solutions of these NLEEs model the dynamics of shallow waters waves near ocean beaches, and in lakes and rivers. The Korteweg-de Vries [KdV] equation that models shallow water waves was very much investigated. Nevertheless, the related Boussinesq equation provides a superior approximation to the adequate description of such water waves.

%%%%%%
In 1871, Boussinesq  proposed the following equation
\begin{equation}\label{Y1}
\begin{aligned}
U_{tt}-U_{xx}+\beta(U^{2})_{xx}+\gamma  U_{xxxx}=0,
\end{aligned}
\end{equation}
where $\beta$ and $\gamma$ are arbitrary constants. This NLEE is completely integrable, possesses an infinite number of conservation laws and has its Lie group symmetries. The Boussinesq equation describes the propagation of shallow water waves with small amplitudes as they propagate at a uniform speed in a water canal of constant depth and it arises in several other physical contexts. Several methods have be used to deal with eq. \eqref{Y1}, including the Lie group method, the inverse scattering transform [IST], the Darboux transformation [DT], the Hirota bilinear method \cite{zz3,pa1,ev1,ev2,ev3, ev4,ev5}, see also diverse physical applications, including one-dimensional nonlinear lattice waves \cite{mt1}, ion sound waves in plasma \cite{mt2}, and vibrations of a nonlinear string \cite{ev4}.  Note that very recently, Clarkson and Dowie studied the rational solutions of the Boussinesq equation and application to RWs and obtained families of rational solutions of the KPI equation from the rational solutions of eq. \eqref{Y1} \cite{mt3}.

Recently, Zhu \cite{jy1} has found a new integrable $(2+1)d$  Boussinesq equation
\begin{equation}\label{N1}
\begin{aligned}
U_{xx}+\alpha U_{yt}-\alpha U_{yy}+\alpha_{1} \epsilon U_{xy}+\alpha_2 \epsilon (U^{2})_{xx}+\alpha_3 \epsilon^{2} U_{xxxx}=0,\; \epsilon^2=\pm 1,
\end{aligned}
\end{equation}
where $U=U(x,y,t)$ is a differentiable function and $\alpha, \alpha_1, \alpha_2, \alpha_3$ are arbitrary constants. Note that for the sake of generality we have inserted four arbitrary parameters in the original equation studied by Zhu \cite{jy1}. The line-type solitons and rational solutions obtained in Ref. \cite{jy1} depend on different choices of the kernel of the Dbar problem.

In this paper, we focus on the exact families of solutions of the more general equation \eqref{N1}.
We derive one-soliton solutions of both bright and dark types for the extended $(2+1)d$ Boussinesq equation \eqref{N1}
by using the traveling wave method. The effects of free parameters on the obtained one-soliton solutions are also analyzed.
We obtain the explicit Hirota bilinear form and the general $N$-soliton solutions, breather, and rational solutions are given.
 An  effective tool, namely the contour line method is applied to study the localization characteristics of the profiles of the obtained RWs \cite{yf1,PD4,yf3,yf4,yf5}. It is well known that the contour line above the asymptotic plane is a closed curve, while the contour line on the asymptotic plane is a hyperbola. A simple question arises: what is the form of the contour line of the obtained lump solution? We thus also study in this paper the profile of the lump solution  by using the
 contour line method. Moreover, we show that rational and semi-rational solutions can be generated by taking the corresponding long wave limit.

The organization of this paper is as follows. In sect. 2, two generic types of one-soliton solutions are constructed by using the traveling wave method. In sect. 3, breathers and RWs were derived by employing the Hirota bilinear method. In sect. 4, semi-rational solutions are generated by taking the corresponding variant of the long wave limit. The main results of this paper are summarized in sect. 5.

\section{One-soliton solutions of the extended $(2+1)d$  Boussinesq equation} \label{1}

In this section, we will derive the analytical one-soliton solutions of the extended $(2+1)d$ Boussinesq equation \eqref{N1} by employing the traveling wave method. The effects of the variation of the parameters $\alpha, \alpha_1, \alpha_2, \alpha_3$ on the field profiles of one-soliton solutions of both bright and dark types are also illustrated by corresponding plots. In this paper,
we set the parameter $\epsilon=1$ for all specific examples and figures.

\subsection{The bright one-soliton solutions}

The solitary wave is regarded as a localized wave of translation that arises from the balance between nonlinear and dispersive or diffractive effects. In order to find analytically the bright one-soliton solution, we consider the following Ansatz
\begin{equation}\label{sN1}
\begin{aligned}
U(x,y,t)=Asech^{p}\{\eta_{1}x+\eta_{2}y-vt\},
\end{aligned}
\end{equation}
where $A,\eta_1,\eta_2$, and $v$ are, respectively, the amplitude, the inverse width, and the velocity of the soliton. The unknown index $p$, where $p>0$, will be determined later.

Substituting eq. \eqref{sN1} into eq. \eqref{N1} and equating the highest exponents of $sech^{p+4}\theta$ and $sech^{2p+2}\theta$ functions, one gets
\begin{equation}\label{sN3}
\begin{aligned}
2p+2=p+4,
\end{aligned}
\end{equation}
which gives $p=2$, where $\theta=\eta_{1}x+\eta_{2}y-vt$.

Setting the coefficients of the same exponent of $sech^{p+j}\theta$ to zero, where $j=0, 2, 4$, since these functions are linearly independent, we obtain a set of algebraic equations:
\begin{equation}\label{sN4}
\begin{aligned}
&v^2-\alpha\eta_2v-\alpha\eta_2^2+\alpha_1\epsilon\eta_1\eta_2+4\alpha_3\epsilon\eta_1^4=0,
\end{aligned}
\end{equation}
\begin{equation}\label{sN5}
\begin{aligned}
&-v^2+\alpha\eta_2+\alpha\eta_2^2-\alpha_1\epsilon\eta_1\eta_2-4\alpha_2\epsilon A^2\eta_1^2+20\alpha_3\epsilon\eta_1^4=0,
\end{aligned}
\end{equation}
\begin{equation}\label{sN6}
\begin{aligned}
&-20A^2\alpha_2 \epsilon\eta_1^2+120A\eta_1^4\alpha_3\epsilon^2=0.
\end{aligned}
\end{equation}
Solving eq. \eqref{sN4}, we obtain
\begin{equation}\label{sN7}
\begin{aligned}
&v=\frac{1}{2}\{\alpha\eta_2\pm \sqrt{((\alpha^2+4\alpha)\eta_2^2-4\alpha_1\epsilon\eta_1\eta_2-16\epsilon^2\eta_1^4\alpha3)}\} ,\\
\end{aligned}
\end{equation}
From eq. \eqref{sN6}, we obtain
\begin{equation}\label{sN8}
\begin{aligned}
&A={\frac {{6\eta_1}^{2}\alpha_3\,\epsilon}{\alpha_2}}.
\end{aligned}
\end{equation}

Substituting $A$ and $v$ into eq. \eqref{sN5}, we see that \eqref{sN5} is automatically satisfied, hence eq. \eqref{sN5} can be understood as the integrability condition for the solitary wave solution. Now eq. \eqref{sN8} imposes the constraint condition  on the parameters as $\alpha_2\alpha_3>0$, for the soliton to exist, and  eq. \eqref{sN7} shows that the velocity $v$ is dependent on $\alpha, \alpha_1$, and $\alpha_3$, and $\pm$ sign in the soliton velocity shows the moving direction of solitary waves. Finally, the above expressions of $A$ and $v$  are substituted in eq. \eqref{sN1}, and we get the solitary wave solution for the extended $(2+1)d$ Boussinesq equation as
\begin{equation}\label{sN9}
\begin{aligned}
U={\frac {{6\eta_1}^{2}\alpha_3\,\epsilon}{\alpha_2}}sech^2\{\eta_{1}x+\eta_{2}y-vt\}.
\end{aligned}
\end{equation}

\begin{figure}[!htbp]
\centering
\subfigure[t=0]{\includegraphics[height=5cm,width=5.5cm]{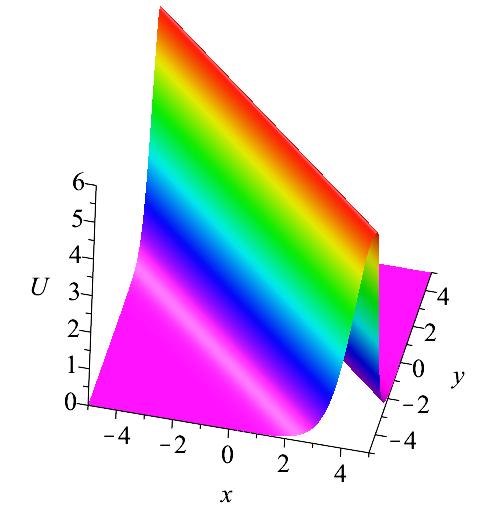}}
\subfigure[t=0]{\includegraphics[height=5cm,width=5.5cm]{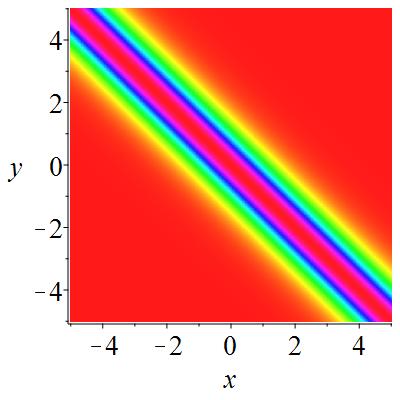}}
\caption{The bright one-soliton solution given by eq. \eqref{sN9} with parameters  $\alpha=1, \alpha_1=-1, \alpha_2=-1, \alpha_3=-1, \eta_1=1, \eta_2=1.$} \label{fig1}
\end{figure}

Figure \ref{fig1} shows the profile of the bright one-soliton solution of the $(2+1)d$ generalized Boussinesq equation. The velocity of the solitary wave is equal to $3$, for our choice of the free parameters.

We briefly analyze the effects of the free parameters of the obtained bright one-soliton solution on both its amplitude and the two widths in the $x$ and $y$ directions. From the expression of the  soliton amplitude $A$ we see that if we fix the parameter $\epsilon$ the soliton amplitude is directly proportional to $\eta_1^{2}$ and $\alpha_3$, and is inverse
proportional to the parameter $\alpha_2$. Thus the amplitude remains unchanged if we vary the soliton parameter $\eta_2$. Also, the two soliton widths in the $x$ and $y$ directions are directly proportional to $\eta_1^{-1}$ and  $\eta_2^{-1}$, respectively. Thus, if we vary $\alpha_2$ and $\alpha_3$ the coresponding soliton widths in the $x$ and $y$ directions remain unchanged.

\subsection{The dark one-soliton solutions}

In order to seek dark one-soliton solutions of eq. \eqref{N1}, the preliminary assumption is
\begin{equation}\label{dN1}
\begin{aligned}
U(x,y,t)=A\tanh^{p}\{\eta_{1}x+\eta_{2}y-vt\},
\end{aligned}
\end{equation}
where $A, \eta_1$, and $\eta_2$ are free unknown parameters and $v$ is the velocity of the soliton. Additionally, the values of the unknown exponents $p$ will be determined later.

Substituting eq. \eqref{dN1} into \eqref{N1}, by equating the highest exponents of $\tanh^{p+4}\theta$ and $\tanh^{2p+2}\theta$ functions, one obtains $p=2$. Collecting the coefficients of the same exponent of $\tanh^{p}\theta$, where  $\theta=\eta_{1}x+\eta_{2}y-vt$, we obtain the following system of algebraic equations:
\begin{equation}\label{dN4}
\begin{aligned}
&v^2-\alpha\eta_2v-\alpha\eta_2^2+\alpha_1\epsilon\eta_1\eta_2-8\alpha_3\epsilon\eta_1^4=0,
\end{aligned}
\end{equation}
\begin{equation}\label{dN5}
\begin{aligned}
&-2v^2+2\alpha\eta_2+2\alpha\eta_2^2-2\alpha_1\epsilon\eta_1\eta_2+3\alpha_2\epsilon A^2\eta_1^2+34\alpha_3\epsilon\eta_1^4=0,
\end{aligned}
\end{equation}
\begin{equation}\label{dN6}
\begin{aligned}
&3v^2-3\alpha\eta_2-3\alpha\eta_2^2+3\alpha_1\epsilon\eta_1\eta_2-16\alpha_2\epsilon A^2\eta_1^2-120\alpha_3\epsilon\eta_1^4=0,
\end{aligned}
\end{equation}
\begin{equation}\label{dN7}
\begin{aligned}
&20A^2\alpha_2 \epsilon\eta_1^2+120A\eta_1^4\alpha_3\epsilon^2=0.
\end{aligned}
\end{equation}
From eq. \eqref{dN4}, the soliton velocity is determined as
\begin{equation}\label{dN8}
\begin{aligned}
v=\frac{1}{2}\{\alpha\eta_2\pm \sqrt{((\alpha^2+4\alpha)\eta_2^2-4\alpha_1\epsilon\eta_1\eta_2+32\epsilon^2\eta_1^4\alpha_3)}\},
\end{aligned}
\end{equation}
and eq. \eqref{dN7} leads to
 \begin{equation}\label{dN9}
\begin{aligned}
A={-\frac {{6\eta_1}^{2}\alpha_3\,\epsilon}{\alpha_2}}.
\end{aligned}
\end{equation}

Note that by substituting the value of $v$ and $A$ in eqs. \eqref{dN8} and \eqref{dN9}, we will find that eqs. \eqref{dN5} and \eqref{dN6} are automatically satisfied. Accordingly, these two equations can be understood as the integrability conditions for the domain-wall solutions (dark-type solutions) to exist. The constraint condition is given by $\alpha_2\alpha_3<0$, which must hold for this type of solutions to exist. Finally, the solution $U$ can be expressed as follows
\begin{equation}\label{dN10}
\begin{aligned}
U={-\frac {{6\eta_1}^{2}\alpha_3\,\epsilon}{\alpha_2}}\tanh^2\{\eta_{1}x+\eta_{2}y-vt\}.
\end{aligned}
\end{equation}

\begin{figure}[!htbp]
\centering
\subfigure[t=0]{\includegraphics[height=5cm,width=5.5cm]{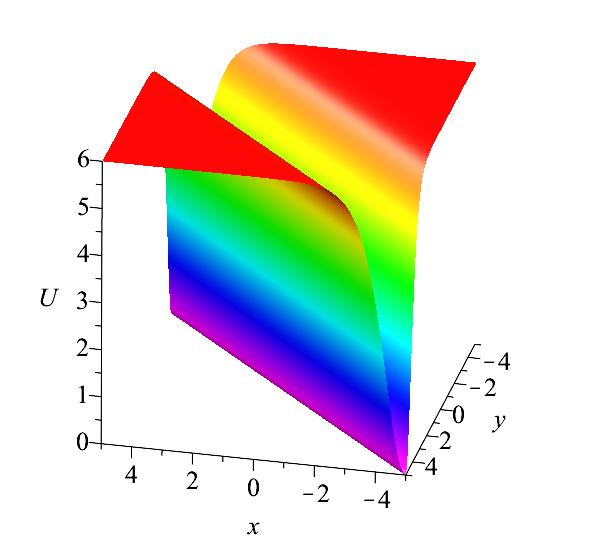}}
\subfigure[t=0]{\includegraphics[height=5cm,width=5.5cm]{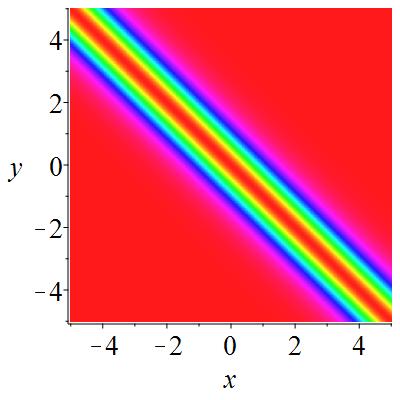}}
\caption{The dark one-soliton solution given by eq. \eqref{dN10} with parameters $\alpha=1, \alpha_1=-1, \alpha_2=-1, \alpha_3=1, \eta_1=1, \eta_2=1.$  }\label{fig2}
\end{figure}

From the above results, we observe that the existence conditions for bright one-soliton solution and dark one-soliton solution  are opposite to each other. Figure \ref{fig2} shows the wave profile of dark one-soliton solution \eqref{dN10}, which satisfies the associated constraint condition for its parameters.

We also briefly analyze the effects of the free parameters of the obtained dark one-soliton solution on both its amplitude and the two widths in the $x$ and $y$ directions. From the expression of the soliton amplitude $A$ we see that if we fix the parameter $\epsilon$ the soliton amplitude is directly proportional to $\eta_1^{2}$ and $|\alpha_3|$, and is inverse
proportional to the parameter $|\alpha_2|$. Thus the amplitude remains unchanged if we vary the soliton parameter $\eta_2$. Also, the two soliton widths in the $x$ and $y$ directions are directly proportional to $\eta_1^{-1}$ and  $\eta_2^{-1}$, respectively. Thus, if we vary $\alpha_2$ and $\alpha_3$ the coresponding soliton widths in the $x$ and $y$ directions remain unchanged.

\section{Breather and rogue wave solutions of the extended $(2+1)d$ Boussinesq equation}\label{2}

In this section, we will derive breather and RW solutions of eq. \eqref{N1} by using the Hirota bilinear method and the long wave limit. Euation \eqref{N1} can be transformed into the bilinear form
\begin{equation}\label{B1}
\begin{aligned}
&(D^{2}_{t}+\alpha D_{y}D_{t}-\alpha D^{2}_{y}+\alpha_1\epsilon D_{x}D_{y}+\alpha_3\epsilon^2D^{4}_{x})f \cdot f =0,\\
\end{aligned}
\end{equation}
through the dependent variable transformation:
\begin{equation}\label{B2}
\begin{aligned}
U=\frac{6\alpha_3 \epsilon}{\alpha_2}(logf)_{xx},\\
\end{aligned}
\end{equation}
where $f$ is a real function, and $D$ is the Hirota bilinear differential operator.

\begin{figure}[tbh]
\centering
\subfigure[t=0]{\includegraphics[height=4cm,width=5.5cm]{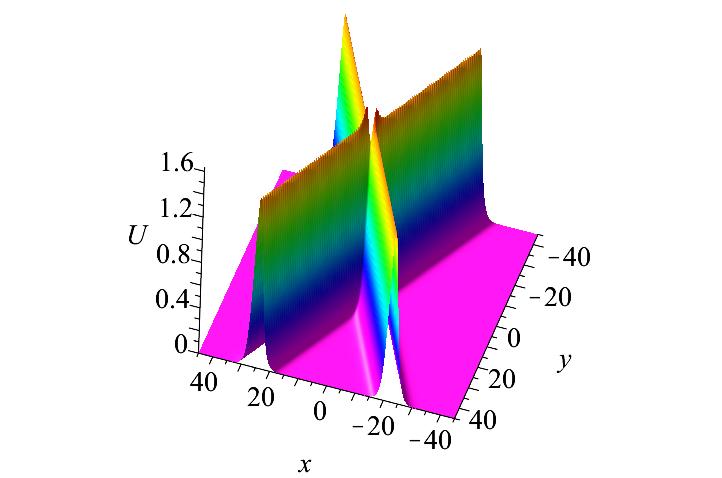}} %
\subfigure[t=0]{\includegraphics[height=4cm,width=5.5cm]{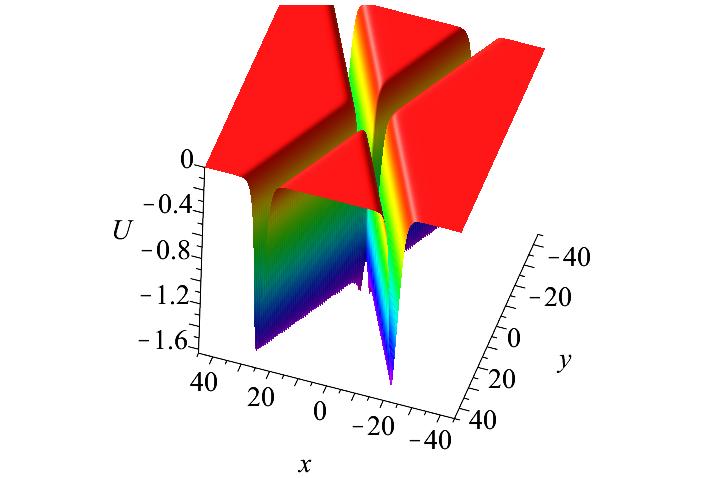}} %
\caption{Two-soliton solutions of eq.(\ref{N1}) in the $(x,y)$-plane at $t=0$, (a): $\alpha_2=1$, (b): $\alpha_2=-1$.}
\label{figs1}
\end{figure}

The $N$-soliton solutions $U$  given in \eqref{B2} of the  extended  $(2+1)d$ Boussinesq equation can be obtained by the bilinear transform method \cite{hirota}, in which $f$ is written in the following form:
 \begin{equation}\label{B3}
\begin{aligned}
f=&\sum_{\mu=0,1}\exp(\sum_{j<k}^{(N)}\mu_{j}\mu_{k}A_{jk}+\sum_{j=1}^{N}\mu_{j}\eta_{j}).
\end{aligned}
\end{equation}
Here
\begin{equation}\label{B4}
\begin{aligned}
&\eta_{j}=P_{j}x+Q_{j}y+\Omega_{j}t+\eta^{0}_{j},\\
&\Omega_{j}=-\frac{1}{2}\{\alpha Q_j-\sqrt{((\alpha^2+4\alpha)Q_j^2-4\alpha_1\epsilon P_jQ_j-4\epsilon^2P_j^4\alpha_3)}\},\\
&A_{jk}=\frac{-\alpha_2 A^4\epsilon^2-\alpha_1AB\epsilon+B\alpha(B-C)-C^2}{\alpha_2 A_*^4\epsilon^2+\alpha_1A_*B_*\epsilon-B_*\alpha(B_*-C_*)+C_*^2},\\
&A=P_j-P_k, A_*=P_j+P_k, B=Q_j-Q_k, B_*=Q_j+Q_k, C=\omega_j-\omega_k, \\
&C_*=\omega_j+\omega_k,
\end{aligned}
\end{equation}
where $P_{j}, Q_{j}, \eta^{0}_{j}$ are arbitrary real parameters,
 and the subscript $j$ denotes an integer. The notation $\sum_{\mu=0}$ indicates summation over all possible combinations of $\mu_{1}=0,1,\mu_{2}=0,1,\cdots,\mu_{n}=0,1$.  The $\sum\limits_{j<k}^{(N)}$ summation is over all possible combinations of the $N$ elements with the specific condition $j<k$. We must emphasize that  $(\alpha^2+4\alpha)Q_j^2-4\alpha_1\epsilon P_jQ_j-4\epsilon^2P_j^4\alpha_3)>0$ must hold.
 For simplicity, the one-soliton solution $U_{soliton}$ can also be expressed in terms of hyperbolic function as
 $$ U_{soliton}=3\,{\frac {\alpha_3\,\epsilon\,{{\it P_1}}^2{\rm sech} \left(\tilde{\theta}
\right)}{\alpha_2\, \left[ 1+{\rm sech} \left(\tilde{\theta}\right) \right]^2 }}, $$
where $\tilde{\theta}=\frac{1}{2}\{\alpha Q_1+\sqrt{((\alpha^2+4\alpha)Q_1^2-4\alpha_1\epsilon P_1Q_1-4\epsilon^2P_1^4\alpha_3)}\}-P_1x-Q_1y-\eta^0_1$. It is  not difficult to find that the $U_{soliton}$ is equivalent to eq. \eqref{sN9} by choosing suitable parameters. Taking $N=2, \alpha_2=-1,\alpha=-1,\alpha_1=1, \alpha_3=-1,\eta^0_1=\eta^0_2=0,P_1=P_2=1,Q_1=-Q_2=1/2$, the two-soliton solutions can be generated by eqs. (\ref{B3}) and (\ref{B4}). These solutions are plotted in Fig. \ref{figs1}.

 Following our earlier works \cite{rao1,rao2}, the $n$th-order breather solutions of the
extended $(2+1)$-dimensional Boussinesq equation can be generated by taking the set of parameters in eq.\eqref{B3}
\begin{equation}\label{B5}
\begin{aligned}
N=2n,P_{j}^{*}=P_{j+1},Q_{j}^{*}=Q_{j+1},\alpha=-1,\alpha_1=1,\alpha_2=-1,\alpha_3=-1.
\end{aligned}
\end{equation}
For example, we take
\begin{equation}\label{B6}
\begin{aligned}
N=2, P_{1}^*=P_{2},Q_{1}^*=Q_{2},
\end{aligned}
\end{equation}

\begin{figure}[tbh]
\centering
\subfigure[t=-8]{\includegraphics[height=2.5cm,width=3.8cm]{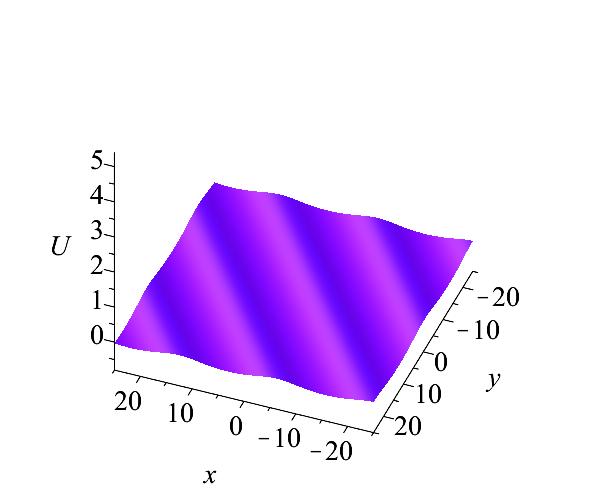}} %
\subfigure[t=-2]{\includegraphics[height=2.5cm,width=3.8cm]{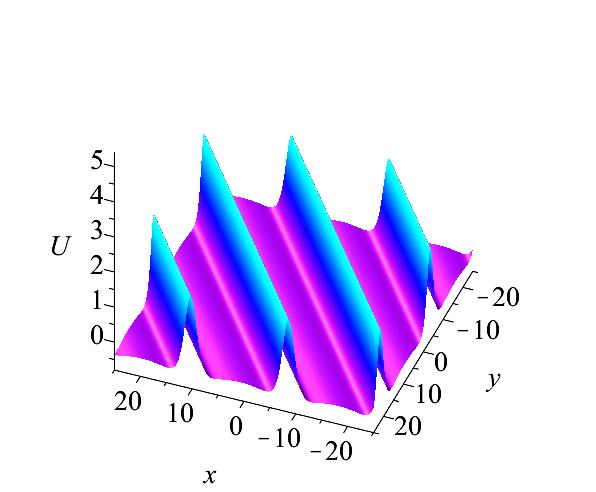}} %
\subfigure[t=0]{\includegraphics[height=2.5cm,width=3.8cm]{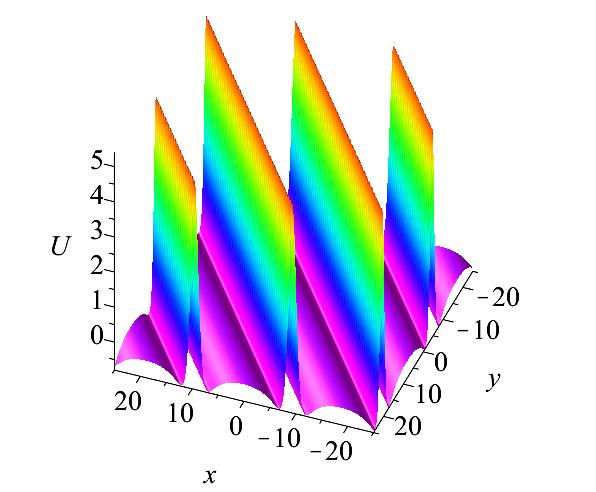}} %
\subfigure[t=0.5]{\includegraphics[height=2.5cm,width=3.8cm]{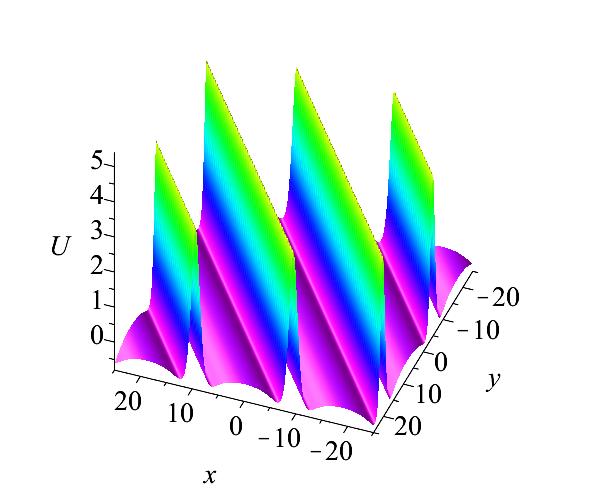}} %
\subfigure[t=8]{\includegraphics[height=2.5cm,width=3.8cm]{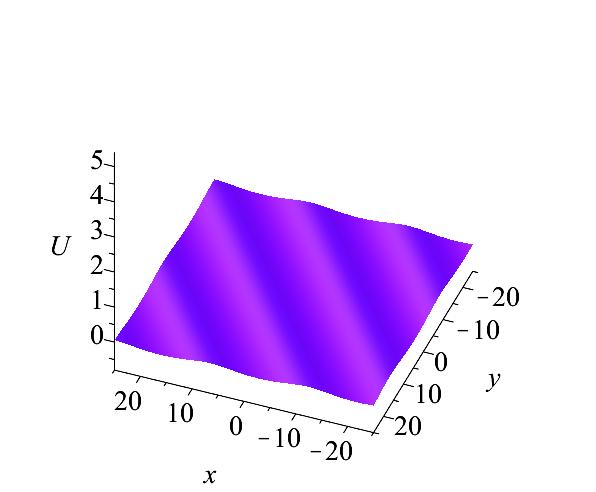}}
\subfigure[]{\includegraphics[height=2.5cm,width=3.8cm]{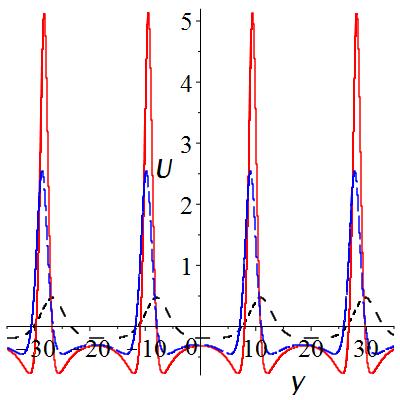}}
\caption{The evolution of the first-order line breather in the $(x,y)$-plane, corresponding to parameters $%
P_{1}=\frac{i}{2},Q_{1}=\frac{i}{3}$, and $\protect\eta _{1}^{0}=0$ in eq. (\protect\ref{B6}),
(f): $t=-4$ (dash, \, black), $0$ (solid, \, red), $1$(longdash, \, blue).}
\label{fig5}
\end{figure}

The first-order breather solution $U$ is derived analytically, and its profile is shown in Fig. \ref{fig5}. As can be seen from  Fig. \ref{fig5}, the first-order breather describes growing and decaying periodic line waves in the $(x,y)$ plane. When $t\ll 0$, these solutions go to a uniform constant background. In the intermediate times, periodic line waves arise from the constant background (see the panels at $t=-8$), and then they attain much higher amplitudes (see the panel at $t=0$). At a larger time, these periodic line waves go back to the constant background (see the panel at $t=8$).

We also give the second-order breather solution. In this case we set the parameters in eq.\eqref{B3}  as follows:
\begin{equation}\label{B7}
\begin{aligned}
N=4, P_{1}^*=P_{2},Q_{1}^*=Q_{2},P_{3}^*=P_{4},Q_{3}^*=Q_{4},\eta _{1}^{0}=\eta _{2}^{0}=\eta _{3}^{0}=\eta _{4}^{0}=0.
\end{aligned}
\end{equation}

\begin{figure}[tbh]
\centering
\subfigure[t=-15]{\includegraphics[height=3cm,width=3.8cm]{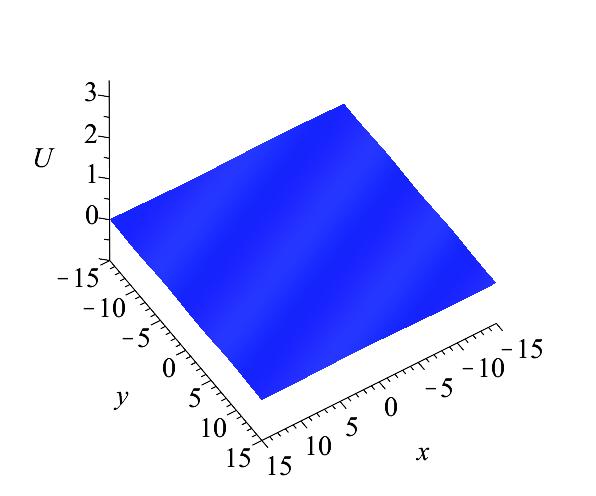}} %
\subfigure[t=-5]{\includegraphics[height=3cm,width=3.8cm]{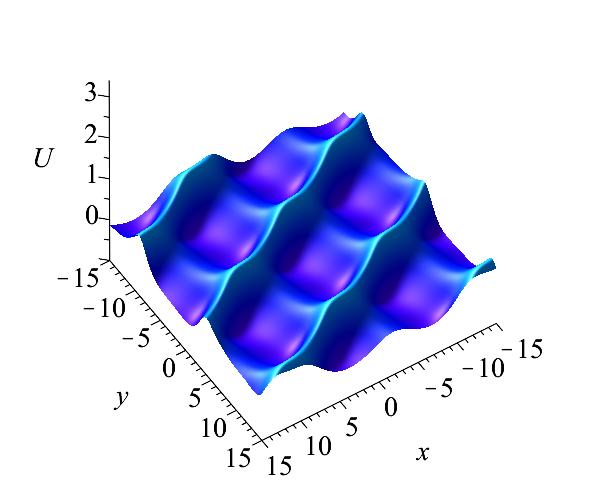}} %
\subfigure[t=-1.25]{\includegraphics[height=3cm,width=3.8cm]{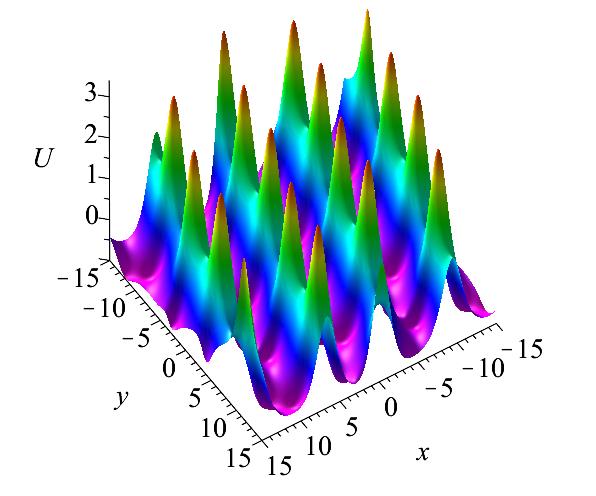}} %
\subfigure[t=1]{\includegraphics[height=3cm,width=4.5cm]{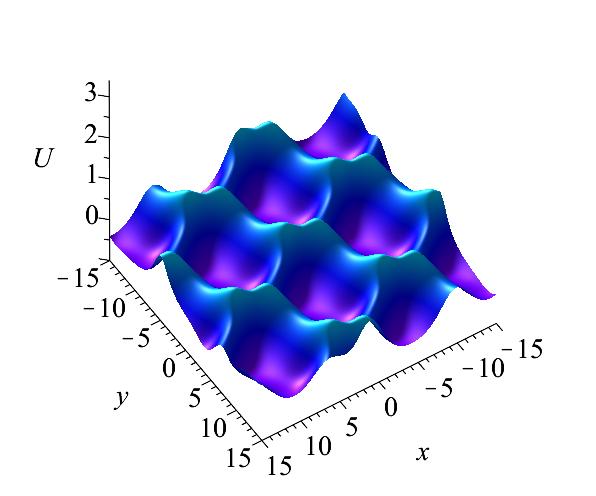}} %
\subfigure[t=10]{\includegraphics[height=3cm,width=4.5cm]{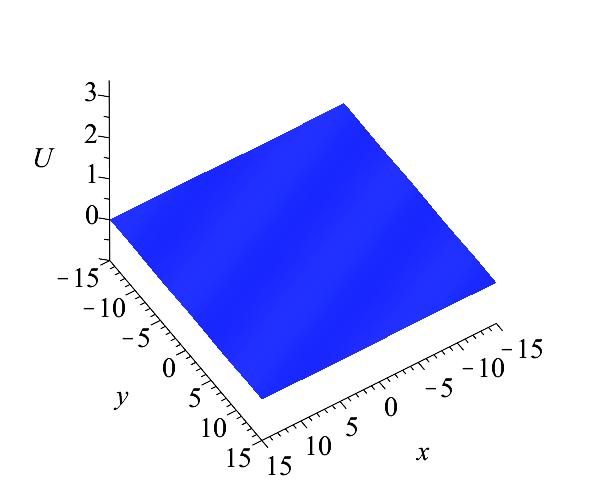}}
\caption{The evolution of the second-order breather solution under conditions in eq. (\ref{B7}) and parameters
$P_{1}=\frac{i}{4}, Q_{1}=-\frac{i}{2}, P_{3}=\frac{i}{2}, Q_3=\frac{i}{4}$.}
\label{fig6}
\end{figure}

\begin{figure}[tbh]
\centering
\subfigure[t=0]{\includegraphics[height=4cm,width=5.5cm]{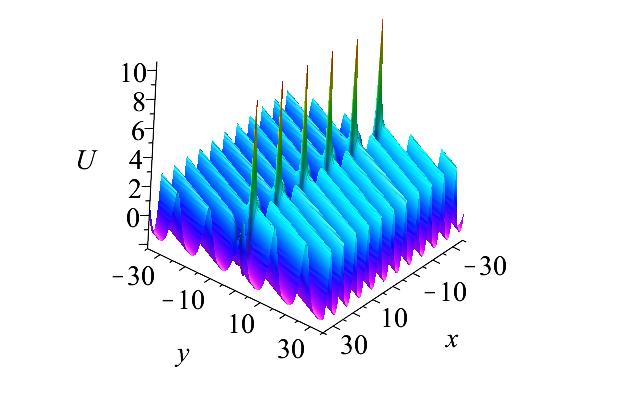}} %
\subfigure[t=0]{\includegraphics[height=4cm,width=5.5cm]{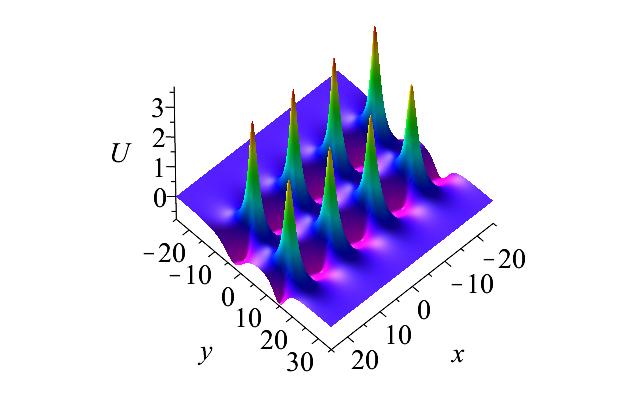}} %
\caption{The second-order breather solution $U$ of eq. (\protect\ref{N1}) in the
$(x,y)$-plane at $t=0$ under conditions  in eq. (\ref{B7}) and  parameters (a): $P_{1}=i, Q_{1}=-\frac{i}{2},P_{3}=\frac{i}{2}, Q_3=1$, (b): $P_{1}=\frac{i}{2}, Q_{1}=-\frac{1}{2},P_{3}=\frac{i}{2}, Q_3=\frac{1}{3}$.}
\label{fig7}
\end{figure}

As can be seen in Fig. \ref{fig6}, the two-line breathers arise from the asymptotic constant background and then interact with each other. This solution reaches a much higher amplitude in the region of intersection and interaction of line breathers. Interestingly, the interactions of the two-line breathers generate doubly-periodic line waves (see the panel at $t=-1.25$). When $t\gg10$ these periodic line waves retreat back to the asymptotic constant background, uniformly in the $(x,y)$-plane. What is interesting, the second-order breather solutions have qualitatively different behaviors in the $(x,y)$-plane, if we modify the values of $P$ and $Q$, see Fig. \ref{fig7}.

For larger values of  $N$, similar to the $N$-th soliton solution, the higher-order breathers can be also obtained analytically and they display a richer dynamical behavior. However, we will not going to investigate here this problem.

%%%%%%%%%%%%%%%%%%%%%%%%%%%%%%%%%%%%%%%%%%%%%%%%%%%%%%%%%%%%%%%%%%%%%%%%%%%%%%%%%%%%%%%%%%%%%%%%%%%%%%%%%%%%%%%%%%%%%%%%%%%%%%%%%%%%%%%%%%%%%%%%%%%%%%%%%%%%%%%%%%%%%
 The rational solutions of eq. (\ref{N1}) are generated from breathers given by eq. (\ref{B3}) under the long-wave limit. Taking
the parameters in eq. \eqref{B3} as
\begin{equation}
N=2,Q_{1}=\lambda _{1}P_{1},Q_{2}=\lambda _{2}P_{2},\lambda^* _{1}=\lambda
_{2},  \alpha=\alpha_3=-1, \alpha_1=\alpha_2=1,\label{N9}¡¤
\end{equation}%
and taking the limit of $P_{j}\rightarrow 0$ $(j=1,2)$, the first-order rational
solution is obtained in the following form
\begin{equation}\label{B10}
\begin{aligned}
U=\theta_{1}\theta _{2}+\alpha _{12},
\end{aligned}
\end{equation}
where
\begin{equation}\label{B11}
\begin{aligned}
&\theta _{j}=\frac{1}{2}\{(-3\lambda_j^2-4\lambda_j)^{\frac{1}{2}}+\lambda_j(t+2y)+2x\}, \\
&\alpha _{jk}=24\alpha_3\{(3\lambda_j^2+4\lambda_j)^{\frac{1}{2}}(3\lambda_k^2+4\lambda_k)^{\frac{1}{2}}\gamma_j\gamma_k
+3\lambda_j\lambda_k+2\lambda_j+2\lambda_k\}.
\end{aligned}
\end{equation}

The rational solutions can be classified into two types: RWs and lumps. {The corresponding rational solutions are line RWs, when $\lambda_{k}$ is real \cite{OYY1,OYY2}, and are lumps when $\lambda_{k}$ is complex  \cite{sj1,sj2}}. Then, the following two cases of rational solutions are considered for the extended $(2+1)d$  Boussinesq equation.

\begin{figure}[tbh]
\centering
\subfigure[t=-13]{\includegraphics[height=2.7cm,width=3.8cm]{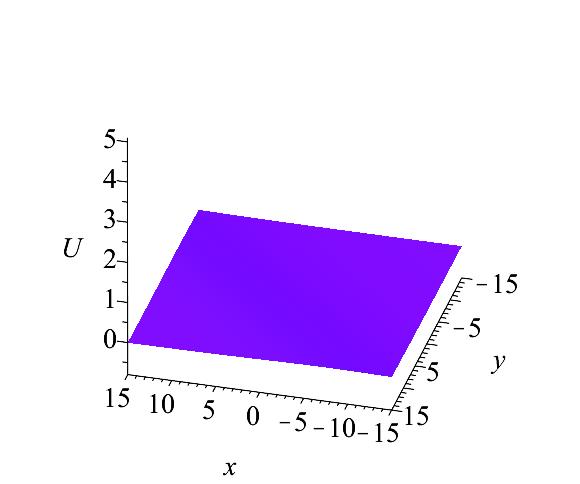}} %
\subfigure[t=-3]{\includegraphics[height=2.7cm,width=3.8cm]{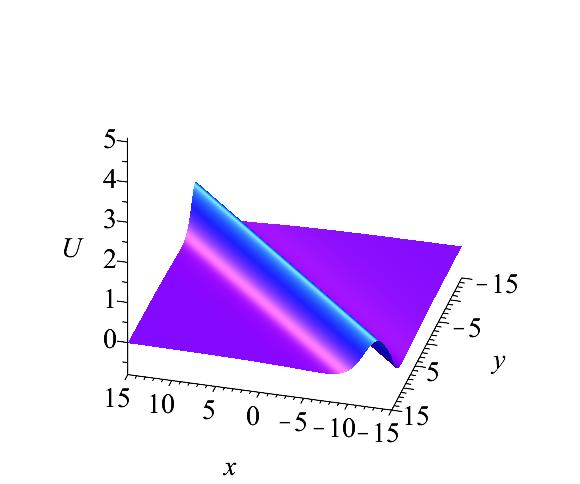}} %
\subfigure[t=0]{\includegraphics[height=2.7cm,width=3.8cm]{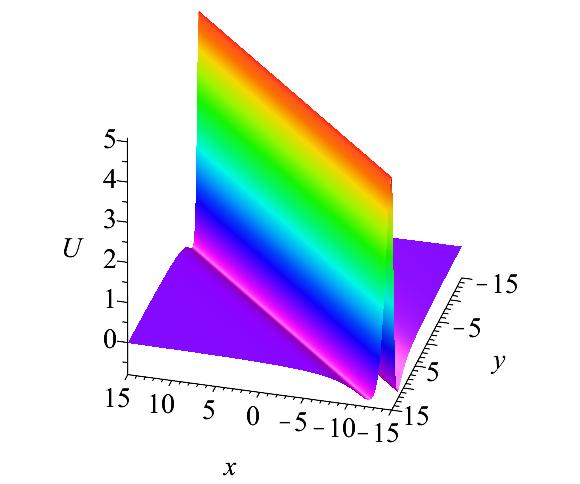}} %
\subfigure[t=1.5]{\includegraphics[height=2.7cm,width=4.5cm]{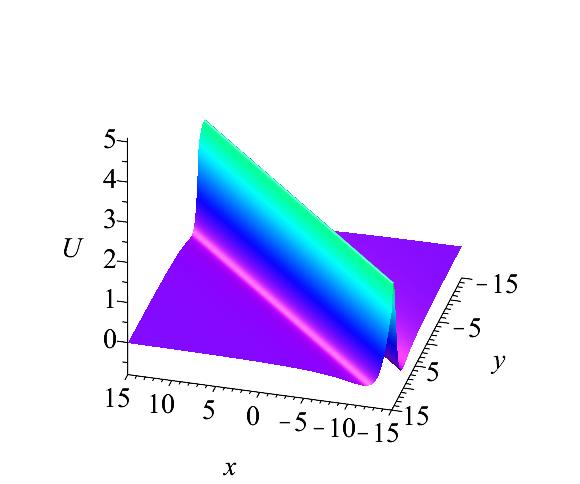}} %
\subfigure[t=13]{\includegraphics[height=2.7cm,width=4.5cm]{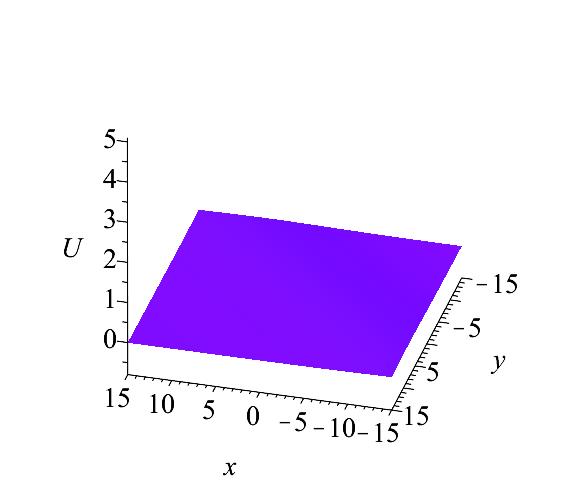}}
\caption{The time evolution in
the $(x,y)$ plane of the first-order (W-shaped) line rogue wave given by solution \eqref{G2}.}
\label{fig9}
\end{figure}

\noindent\textbf{ Case 1 (RWs):} In order to obtain RW solutions from eq. \eqref{B10}, we set
\begin{equation}\label{G1}
\begin{aligned}
\lambda _{1}=\lambda_{2}=1, \gamma_1=-\gamma_2=1.
\end{aligned}
\end{equation}
The RW solution is
\begin{equation}\label{G2}
\begin{aligned}
U_{RW}=\frac{-1200(2x+2y+t)^2+6000t^2+11520}{[5(2x+2y+t)^2+25t^2+48]^2},
\end{aligned}
\end{equation}
and the corresponding profile of RW solution $U_{RW}$ is shown in Fig.
\ref{fig9}. It is seen that this W-shaped solution describes an emerging and
decaying line wave oriented in the $(1, -1)$ direction of the
$(x, y)$-plane. At any given time, this solution keeps a constant value along
the line direction defined by $2y+2x+c=0$, and the solution $U_{RW}$ uniformly approaches the constant background at
$t\rightarrow \pm \infty $. At finite time, $U_{RW}$ attains
the maximum value $5$ (i.e., five times the background amplitude) at the
center ($y+x=0$) of the line wave at $t=0$. {We note that in general, the breather solution is a kind of travelling periodic solution;
see, for example, the second-order breather solution shown in Fig. \ref{fig7}(b), which consists of two
separate breather solitons. In contrast, the rogue wave solution is a kind of rational solution that has a significant
amplitude growth during a short interval of time around $t=0$.  Figure \ref{fig9} clearly shows this
unique feature of the rogue wave solution.}
We stress that the amplitude of the RW is changing,
and it is also moving at a speed equal to  $1$ for our choice of parameters.

\noindent\textbf{ Case 2 (Lump solution):} It  can be obtained from eq.\eqref{B10}  as
\begin{equation}\label{L1}
\begin{aligned}
U_{Lump}=\frac{12(x+t)^2-12y^2-36}{[(x+t)^2+y^2+3]^2}.
\end{aligned}
\end{equation}
Here, the constraint conditions are
\begin{equation}\label{L2}
\begin{aligned}
\lambda _{1}=-\lambda_{2}=i, \gamma_1=\gamma_2=1.
\end{aligned}
\end{equation}

It is easy to see that the lump solution $U_{Lump}$ is constant along the trajectory when $y=0,x+t=0$. If we modify the time variable $t$, the patterns of lump solution do not change. The solution $U_{Lump}$ has the following critical points
\begin{equation}\label{L1a}
\begin{aligned}
A_1=(x_1, y_1)=(0, 0), A_2=(x_2, y_2)=(3, 0), A_3=(x_3, y_3)=(-3, 0).
\end{aligned}
\end{equation}
The lump solution $U_{Lump}$ moves along the direction of $y=0$ and attains the maximum value $4$ at the $(0, 0)$ point and the minimum value $-\frac{1}{2}$ at the $(-3,0)$ and $(3,0)$ points when $t=0$. The solution $U_L$ goes to $0$ when $x\rightarrow \infty$ and $t\rightarrow\infty$, which implies that the hight of the asymptotic background is equal to $0$.

Next we shall use the contour line method applied to eq. \eqref{L1} in order to study the lump profile by  varying the height $d$ of contour line. Using this method, a contour line of eq. \eqref{L1} at height $d$ along the
$(x, y)$-plane is expressed by
{\small
\begin{equation}\label{L2a}
\begin{aligned}
-64(t^2+2tx+x^2+y^2+3)^2d+768t^2+1536tx+768x^2-768y^2-2304=0.
\end{aligned}
\end{equation}
}
Here $d$ denotes the height of contour line from the asymptotic background.

\begin{figure}[tbh]
\centering
\subfigure[]{\includegraphics[height=4.5cm,width=5.5cm]{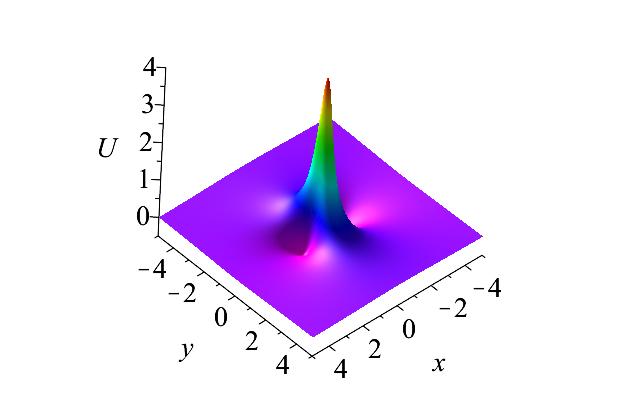}} %
\subfigure[]{\includegraphics[height=4.5cm,width=5.5cm]{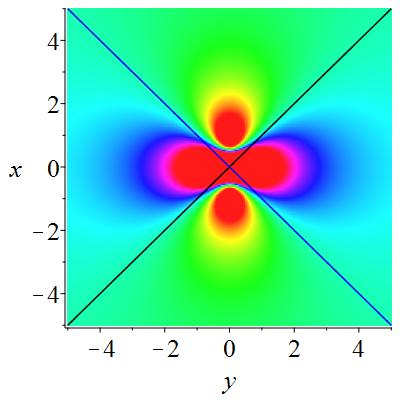}} %  \\
\subfigure[]{\includegraphics[height=4.5cm,width=5.5cm]{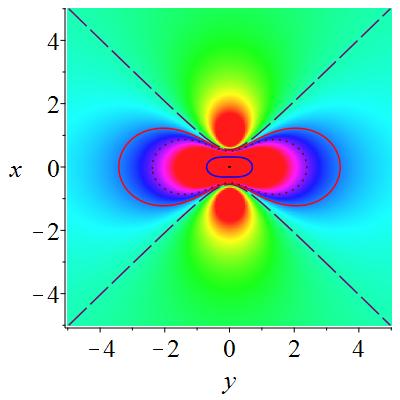}} %
\subfigure[]{\includegraphics[height=4.5cm,width=5.5cm]{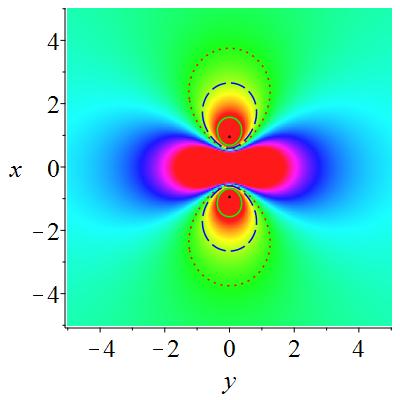}} %
\caption{The lump solution $U_{Lump}$ given by eq. (\protect\ref{L1}), in the $(x,y)$ plane and at $t=0$. }
\label{fig10}
\end{figure}

We set $d=0$ in eq. \eqref{L2a}, and we see that the contour line is a hyperbola on the asymptotic plane, which has two asymptotes:
\begin{equation}\label{L3}
\begin{aligned}
l_1: \, y=x, \,  l_2: \, y=-x.
\end{aligned}
\end{equation}
The two asymptotes are plotted in Fig. \ref{fig10}b. The height of contour line above the asymptotic background must be in the interval $(0,4]$. It is clear from Fig. \ref{fig10}c that the contour line above the asymptotic background is reducing from a single point at height of $4$ to a concave line, and then it becomes a convex line. Finally, it reduces to a hyperbola on the asymptotic plane (for $d=0$).
For the contour line of lump solution below the asymptotic plane, it is not difficult to find that the two separate contour lines are convex for all values of the parameter $d$, and finally they reduce to two separate points (see Fig. \ref{fig10} d).  The two minimum points are located at $(-3,0)$ and $(3,0)$, for our choice of parameters.

The profile of the first-order lump solutions $U_{Lump}$ is shown in Fig. \ref{fig10}a, and the panel $(b)$ is the density plot of the panel $(a)$ of Fig. \ref{fig10}. In panel $(b)$, $l_1$ (longdash, \, blue) and $l_2$ (dash, \, black) are the two asymptotes of the contour line in eq. \eqref{L2a}. Panel $(c)$ shows the contour lines for different values of $d$ from outside to inside: $d=0$ (longdash, yellow), $d=0.1$ (solid,\, red), $d=0.2$ (dash,\, green), $d=1.5$ (solid, \, blue), $d=4$ (a single point). The solution  $U_{Lump}$ reaches the maximum value of $4$ at the origin $(0, 0)$. The parameter $d=0$ produces two branches of a hyperbola on the asymptotic plane. In panel $(d)$ we show the contour lines for different values of the parameter $d$ below the asymptotic background. The contour lines are plotted from the margins of the figure to its center, for $d=-0.08$, $d=-0.15$, $d=-0.34$, and $d=-0.5$, respectively. Two symmetric points with respect to the $x$-axis are clearly visible in the panel $(d)$ for $d=-0.5$.

\section{Semi-rational solutions of the $(2+1)d$ extended Boussinesq equation}\label{3}

Through the above discussion we know that  the rational solutions are generated by taking a long wave limit of all of the exponential functions in $U$. Furthermore, it is natural to derive the semi-rational solutions by taking a long wave limit of only a part of exponential functions in $f$. Setting $0<2j<N$ and $1\leq k\leq 2j$,
\begin{equation} \label{hh1}
\begin{aligned}
Q_{k}=\lambda_{k}P_{k}\,\,,\eta_{k}^{0}=i\pi, \alpha=-1,\alpha_1=1,\alpha_2=-1,\alpha_3=-1,
\end{aligned}
\end{equation}
and taking the limit $P_{k}\rightarrow 0$ for all $k$,  then the functions  $f$ defined in  eq.\eqref{B3} become
 a combination of polynomial  and exponential functions, which generate the semi-rational solutions $U$  of
  the extended $(2+1)d$ Boussinesq equation through eq.\eqref{B2}. Moreover, in order to avoid the singularity of $U$ generated by $f$ and the use of the parameter constraints given in  eq. \eqref{hh1} and the above long wave limit, it is necessary to take the following constraints:
\begin{equation} \label{hh2}
\begin{aligned}
\lambda_{k}=\lambda_{j+k}^{*}\,,\delta_{k}\delta_{j+k}=-1\,(2j+1 \leq s\leq N).
\end{aligned}
\end{equation}

\begin{figure}[tbh]
\centering
\subfigure[t=-4]{\includegraphics[height=3cm,width=3.8cm]{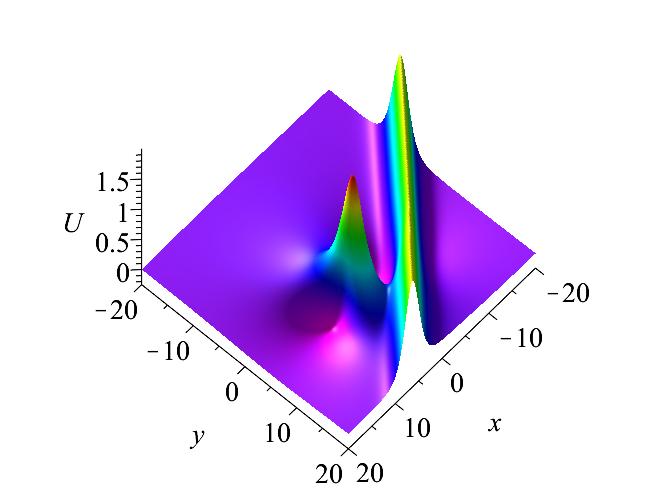}} %
\subfigure[t=0]{\includegraphics[height=3cm,width=3.8cm]{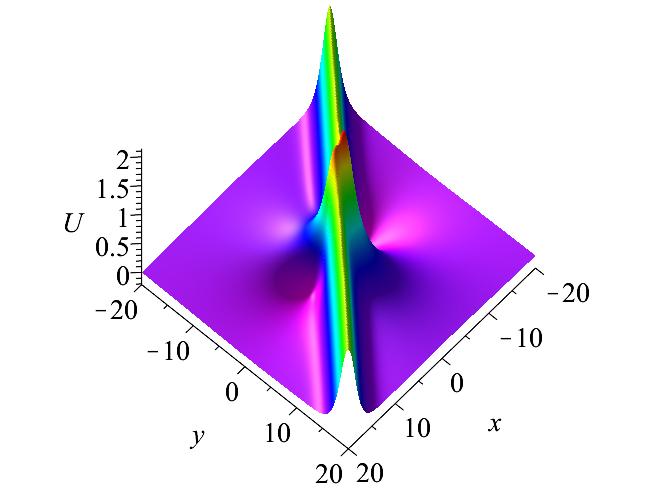}} %
\subfigure[t=4]{\includegraphics[height=3cm,width=3.8cm]{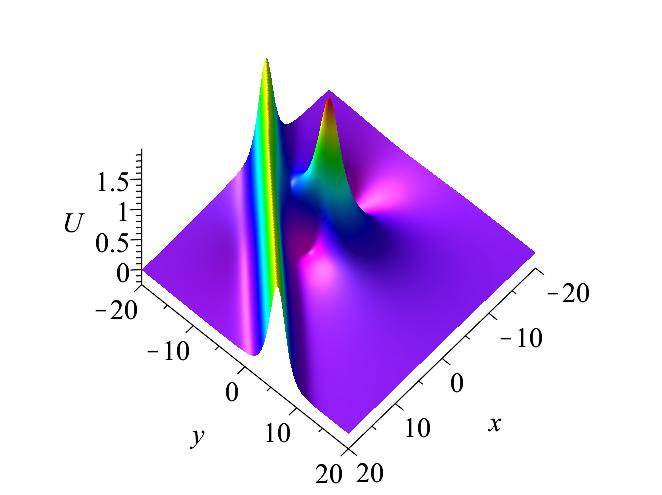}} %
\subfigure[t=-15]{\includegraphics[height=3cm,width=3.8cm]{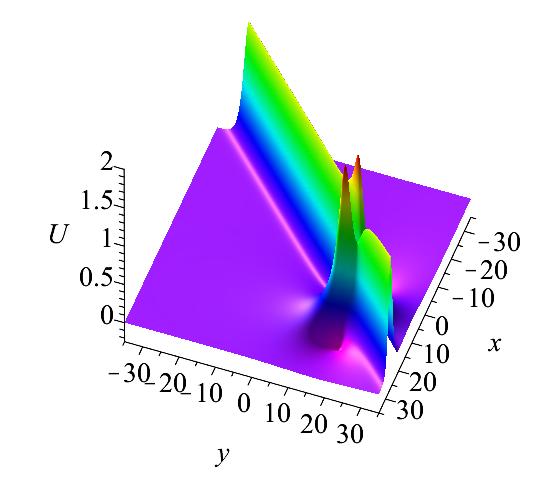}} %
\subfigure[t=0]{\includegraphics[height=3cm,width=3.8cm]{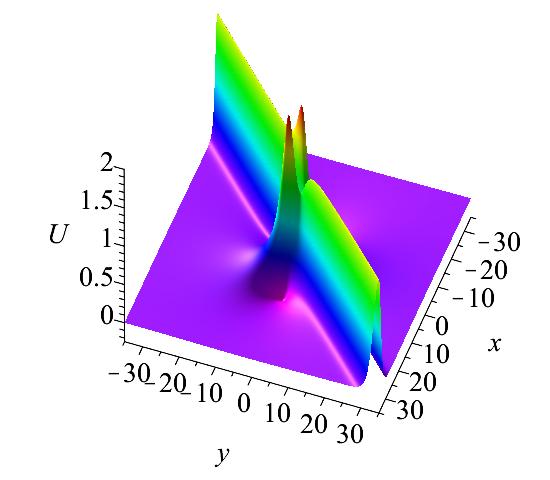}} %
\subfigure[t=15]{\includegraphics[height=3cm,width=3.8cm]{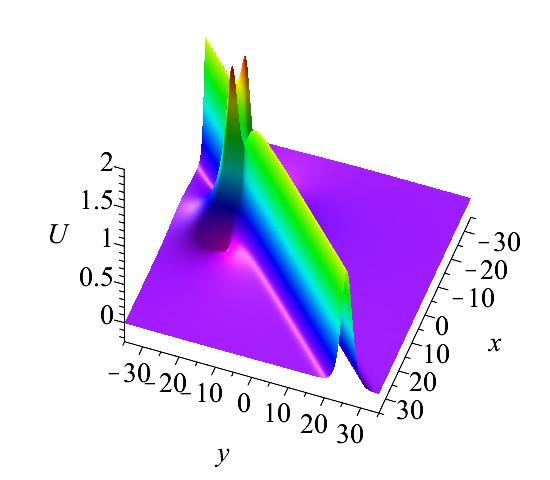}} %
\caption{The time evolution in the $(x,y)$ plane of the two distinct patterns of semi-rational solutions consisting of a
first-order lump and a single-soliton given by eq. (\protect\ref{hh4}), with parameters (a, \, b, \, c): $P_3=-1, \, Q_3=1$; (d, \, e, \, f): $P_3=1, \, Q_3=-1$. }
\label{fig11}
\end{figure}

To clearly illustrate this method for constructing the smooth semi-rational solutions, we first consider the case of $N=3$. Setting
\begin{equation} \label{hh3}
\begin{aligned}
N=3, Q_1=\lambda_1P_1,  Q_2=\lambda_2P_2,\eta_{1}^{0}=\eta_{2}^{0}=i\pi,
\end{aligned}
\end{equation}
and taking $P_1,P_2\rightarrow0$ in eq. \eqref{B3}, we obtain
\begin{equation} \label{hh4}
\begin{aligned}
f=&(\theta_{1}\theta_{2}+a_{12})+(\theta_{1}\theta_{2}+a_{12}+a_{13}\theta_{2}+a_{23}\theta_{1}+a_{12}a_{23})e^{\eta_{3}},\\
\end{aligned}
\end{equation}
where $a_{j3}=12\,{\frac {{{\it P_3}}^{2}}{-4\,{{\it P_3}}^{3}+2\,\sqrt {{\it P_3}\,
 \left( {{\it P_3}}^{3}-{\it Q_3} \right) }\sqrt {-\lambda_j}+\lambda_j\,{
\it P_3}+{\it Q_3}}}
$  and $\theta_j, a_{12}\,,\eta_{3}$ are given by eqs. \eqref{B4} and \eqref{B11}.
The corresponding semi-rational solutions $U$ are hybrids of first-order lump solutions and one-soliton solutions. A generic  type of semi-rational solution is  plotted in Fig. \ref{fig11}. We can see the different patterns of the interaction of lump wave and one-soliton wave, by varying the parameters $P_3$ and $Q_3$. We see that the lump moves and passes the soliton, and in the intersection domain of the two waveforms the amplitude increases considerably, see Fig. \ref{fig11}a, b, c. The lump can also move only along the direction of the peak amplitude of the one-soliton wave, see Fig. \ref{fig11}d, e, f.

Higher-order semi-rational solutions consisting of lump and breather solutions can also be generated in  a similar way for  larger values of $N$. For example,
\begin{equation} \label{hh5}
\begin{aligned}
N=4, Q_1=\lambda_1P_1,  Q_2=\lambda_2P_2,\eta_{1}^{0}=\eta_{2}^{0}=i\pi,
\end{aligned}
\end{equation}
and taking $P_1, P_2\rightarrow0$ in eq. \eqref{B3}, we obtain
\begin{equation} \label{hh6}
\begin{aligned}
f=&e^{A_{34}}(a_{13}a_{23}+a_{13}a_{24}+a_{13}\theta_{2}+a_{14}a_{23}+a_{14}a_{24}+a_{14}\theta_{2}+a_{23}\theta_{1}+a_{24}\theta_{1} \\
&+\theta_{1}\theta_{2}
+a_{12})e^{\eta_{3}+\eta_{4}}+(a_{13}a_{23}+a_{13}\theta_{2}+a_{23}\theta_{1}+\theta_{1}\theta_{2}+a_{12})e^{\eta_{3}} \\
&+(a_{14}a_{24}+a_{14}\theta_{2}+a_{24}\theta_{1}
+\theta_{1}\theta_{2}+a_{12})e^{\eta_{4}}+\theta_{1}\theta_{2}+a_{12}\,,\\
\end{aligned}
\end{equation}
where
$$a_{js}=12\,{\frac {{{\it P_s}}^{2}}{-4\,{{\it P_s}}^{3}+2\,\sqrt {{\it P_s}\,
 \left( {{\it P_s}}^{3}-{\it Q_s} \right) }\sqrt {-\lambda_j}+\lambda_j\,{
\it P_s}+{\it Q_s}}},j=(1,2),$$
 $s=(3,4)$ and $\theta_j$ is defined by eq.\eqref{B11}.
Further, we take
\begin{equation} \label{hh7}
\begin{aligned}
\lambda_{1}=-\lambda_{2}\,,\gamma_{1}\gamma_{2}=-1\,,P_{3}=P^*_{4}\,,Q^*_{3}=Q_{4},\eta^0_3=\eta^0_4.
\end{aligned}
\end{equation}

\begin{figure}[!tbh]
\centering
\subfigure[t=0]{\includegraphics[height=4cm,width=5.5cm]{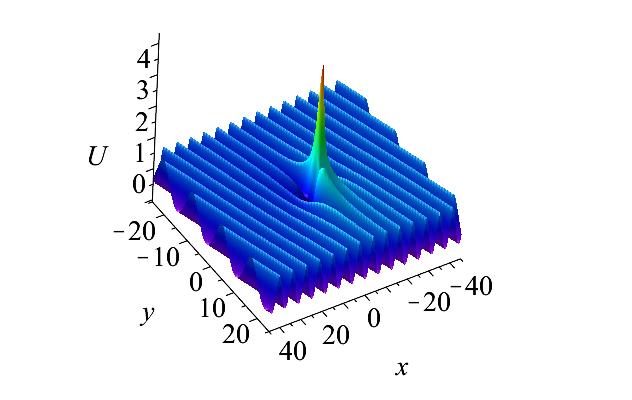}} %
\subfigure[t=0]{\includegraphics[height=4cm,width=5.5cm]{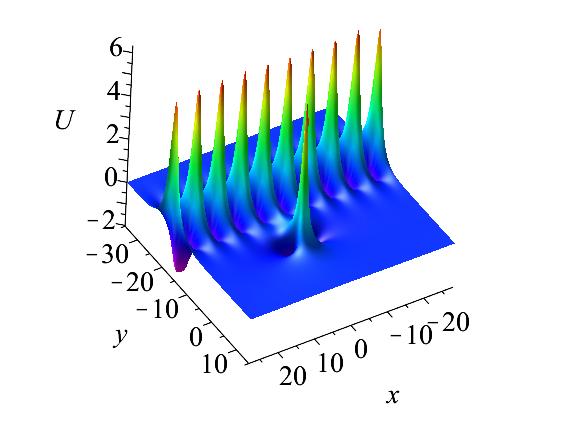}} %
\caption{Semi-rational solutions consisting of a first-order lump and a first-order breather given by eq.\eqref{hh6} at time $t=0$ with parameters (a): $P_{3}=i, \, Q_{3}=-\frac{i}{2}$, (b): $P_{3}=i,\, Q_{3}=\frac{1}{2}$.}
\label{fig12}
\end{figure}

For the above expression of  $f$, semi-rational solutions of  eq.\eqref{B2}, which consist of fundamental lump solution and
one-breather solution are obtained, see Fig. \ref{fig12}. We see from this figure that we get two different kinds of such hybrid solutions: the first one is a mixture of a line breather and a lump (see Fig. \ref{fig12}a), and the second one is the usual breather solution and lump solution (see Fig. \ref{fig12}b).

\section{Summary and discussion}\label{con}

In this paper, the traveling wave method was applied  to construct exact bright and dark one-soliton solutions for the extended $(2+1)$-dimensional Boussinesq equation. The effects of varying the free parameters on the profiles of these exact solutions are also analyzed. The $N$-soliton and $n$th breather  solutions are derived analytically by  the Hirota bilinear method.  The first-order and the second-order breather solutions have been shown (see Fig. \ref{fig5} and Fig. \ref{fig6}). Such solutions constitute arrays of periodic line waves in the $(x, y)$ plane,  and they are growing and then decaying in time. The  W-shaped line RW, the lump solution, and the semi-rational solution have been constructed analytically by taking the corresponding long-wave limit of the above-mentioned periodic solutions.

We studied the localization features of the lump profile by employing the contour line method (see Fig. \ref{fig10}). Moreover, we have shown that several patterns of the derived semi-rational solutions exhibit a range of interesting and rather complicated dynamics (see Fig. \ref{fig11} and Fig. \ref{fig12}). To the best of our knowledge, such type of semi-rational solution composed of a first-order lump and a first-order line breather [see Fig. \ref{fig12} (a)]  has never been reported in the study of other variants of $(2+1)$-dimensional Boussinesq equation. We expect new results in this area helping us to understand both the unique features of nonlinear evolution equations in the multi-dimensional space and the applicability of nonlinear partial differential equations to the description of nonlinear phenomena in diverse physical settings.\\

\noindent\textbf{Conflict statement}. We declare we have no conflict of interests.

\begin{acknowledgements}
This work is supported by the NSF of China under Grant No. 11671219,  and the K.C. Wong Magna Fund in Ningbo University. We thank other members in our group at Ningbo University for many useful discussions on the paper.
\end{acknowledgements}



\end{document}